\documentclass[
 aip,
 long, numerical bibliography, (default)
 jmp,%
 amsmath,amssymb,
preprint,%
nofootinbib]{revtex4-1}

\usepackage[a4paper,margin=2.5cm]{geometry}
\usepackage[utf8]{inputenc}
\usepackage{amsmath}
\usepackage{graphicx}% Include Figure files
\usepackage{dcolumn}% Align table columns on decimal point
\usepackage{bm}% bold math
\usepackage{epstopdf}

\usepackage{pdfpages}

\begin{document}

%\preprint{AIP/123-QED}

%\title{Graphene drum resonators on insulating substrate}
\title{Dynamical strong coupling and parametric amplification in mechanical modes of graphene drums}
\author{John P. Mathew}
\affiliation{Department of Condensed Matter Physics and Materials Science, Tata Institute of Fundamental Research, Homi Bhabha Road, Mumbai 400005, India}
\author{Raj N. Patel}
\affiliation{Department of Condensed Matter Physics and Materials Science, Tata Institute of Fundamental Research, Homi Bhabha Road, Mumbai 400005, India}
\affiliation{Physics Department, Birla Institute of Technology and Science Pilani - K. K. Birla Goa Campus, Goa 403726, India}
\author{Abhinandan Borah}
\author{R. Vijay}
\author{Mandar M. Deshmukh}
\email{deshmukh@tifr.res.in}
\affiliation{Department of Condensed Matter Physics and Materials Science, Tata Institute of Fundamental Research, Homi Bhabha Road, Mumbai 400005, India}

\maketitle

\textbf{Mechanical resonators are ubiquitous in modern information technology. With the ability to couple them to electromagnetic and plasmonic modes, they hold the promise to be the key building blocks in future quantum information technology. Graphene based resonators are of interest for technological applications due to their high resonant frequencies, multiple mechanical modes, and low mass.\cite{bunch_electromechanical_2007,chen_performance_2009,singh_probing_2010,barton_photothermal_2012,singh2014optomechanical,weber_coupling_2014,song_graphene_2014}  The tension mediated non-linear coupling between various modes of the resonator can be excited in a controllable manner.\cite{eriksson2013frequency,westra2010nonlinear,eichler2012strong,castellanos2012strong} Here, we engineer a graphene resonator to have large frequency tunability at low temperatures resulting in large intermodal coupling strength. We observe the emergence of new eigenmodes and amplification of the coupled modes using red and blue parametric excitation respectively. We demonstrate that the dynamical intermodal coupling is tunable.  A cooperativity of 60 between two resonant modes of $\sim$100 MHz is achieved in the strong coupling regime.  The ability to dynamically control the coupling between high frequency eigenmodes of a mechanical system opens up possibility for quantum mechanical experiments at low temperatures.\cite{santamore_quantum_2004,mahboob2014two}}

\begin{figure}[t]
\includegraphics[width=0.70\columnwidth]{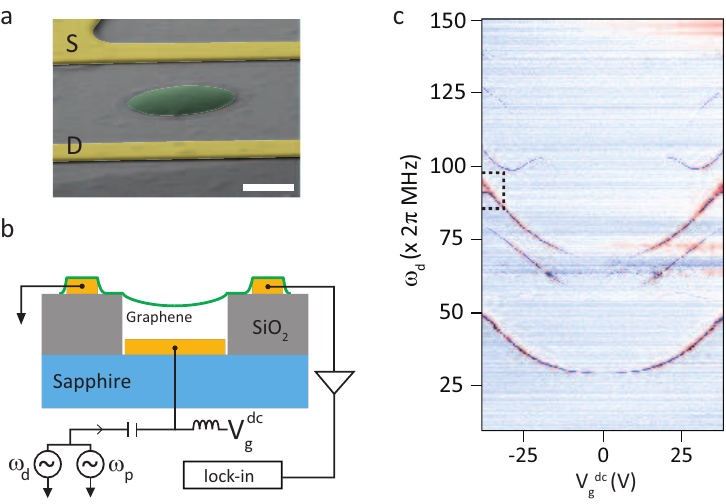}
\caption{\label{schematic} \textbf{Graphene drum electromechanics in the low tension regime.} \textbf{a}, False colored scanning electron microscope image of the graphene drum resonator. The region shaded green shows the suspended part of the graphene. Scale bar is 3 $\mu$m. \textbf{b}, Schematic of the circuit used to actuate and detect the  mechanical modes of the drum. \textbf{c}, Large frequency tunability of the modes with dc gate voltage. The modes near 95 MHz show an avoided crossing (outlined by the black dotted lines).}
\end{figure}
Experiments in cavity optomechanics,\cite{teufel_circuit_2011,massel2011microwave} where a low frequency mechanical oscillator parametrically modulates the resonant frequency of an electromagnetic mode of a cavity, have demonstrated the ability to prepare a mechanical system in its ground state,\cite{wilson-rae_theory_2007,marquardt_quantum_2007,teufel2011sideband,chan2011laser} and entanglement between propagating photons and phonons.\cite{palomaki2013entangling} Since mechanical resonators support a large number of vibrational modes, a natural extension of the optomechanical scheme is to use coupled modes of a mechanical resonator where a high frequency mode plays the role of the cavity.\cite{mahboob2012phonon,okamoto2013coherent} There has been a considerable interest in exploring coupling among eigenmodes of a mechanical system, demonstrating coherent coupling between low frequency modes.\cite{mahboob2012phonon,faust2012nonadiabatic,okamoto2013coherent,faust_coherent_2013} The eventual goal of such experiments is to demonstrate coupling in the quantum regime.\cite{santamore_quantum_2004,mahboob2014two} To achieve a quantum coherent coupling, graphene mechanical resonators offer many advantages. Firstly, the high frequencies of the graphene drum resonators yields lesser phonons once cooled down to low temperatures. Secondly, the large frequency dispersion with gate voltage results in large inter-modal coupling offering advantage while using higher frequency modes as phononic cavity. Furthermore, the large quantum zero-point motion of graphene resonators offers advantage of efficient coupling with electromagnetic cavities.\cite{barton_photothermal_2012,singh2014optomechanical,weber_coupling_2014,song_graphene_2014}

Here we fabricate graphene drum resonators\cite{bunch_electromechanical_2007,chen_performance_2009,singh_probing_2010} by a novel method with low built-in tension yielding large frequency tunability at low temperatures (see Methods for details on fabrication). The actuation and detection of the mechanical modes of graphene is implemented using an all electrical scheme.
The drums we study consist of a graphene flake contacted by Cr/Au electrodes with a central region of diameter 3.5 $\mu$m suspended 300 nm above a local gate electrode of Ti/Pt (Figure \ref{schematic}a,b). The local gate is on a sapphire substrate and the graphene is transferred onto electrodes on SiO$_2$. The region of SiO$_2$ over the gate electrode is etched out prior to placing the flake. The suspended region of the graphene and the gate electrode form an electromechanical system where the capacitance between the gate and graphene is used to transduce the mechanical motion of the drum. The gate electrode serves the dual purpose of actuating the membrane's motion as well as tuning the resonance frequencies of various modes.

\begin{figure}[t]
\includegraphics[width=0.65\columnwidth]{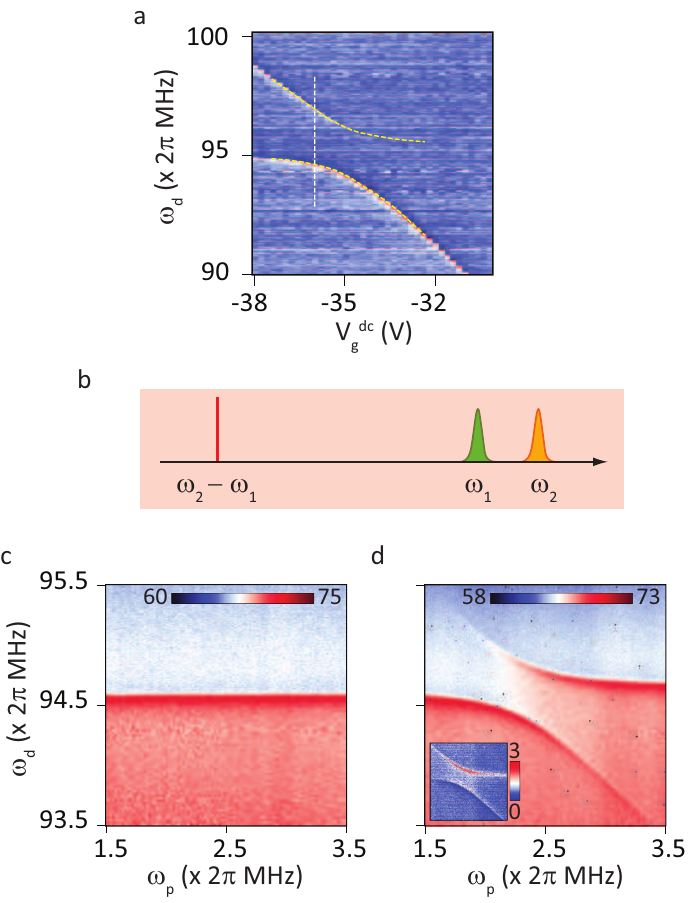}
\caption{\label{moresplits} \textbf{Strong coupling between electromechanical modes.} \textbf{a}, Zoomed in scan of frequency response of the coupled modes as a function of the dc gate voltage showing avoided crossing behavior with no parametric pumping. Red pump experiments are done at the gate voltage indicated by the white dotted line where $V_g^{dc}=-36$ V. \textbf{b}, The red detuned pump is shown alongside the two coupled modes on a frequency axis. Response of mode 1 as a function of the red pump detuning when \textbf{c}, $V_p=0$ V and \textbf{d}, $V_p=1.5$ V. For nonzero pump amplitude, mode 1 is seen to split in the vicinity of $\omega_p\approx\Delta\omega$. Color scale units are $\mu$V. Inset of \textbf{d} shows the response detected at $\omega_d+\omega_p$. Energy transfer to the second mode is seen when $\omega_d+\omega_p\approx\omega_2$.
}
\end{figure}

All measurements were carried out with the sample kept under high vacuum at 5 K in a $^4$He cryostat. Weak radio frequency (\textit{rf}) signals are applied to the gate to drive the resonator,  and the motion is measured using a lock-in detection of the \textit{rf} signal transmitted through the graphene flake. Figure \ref{schematic}b shows the schematic of the circuit used for actuation and detection of the resonator modes. The driving \textit{rf} signal (amplitude: $\tilde{V}_g$, frequency: $\omega_d$)  is combined with an appropriate \textit{rf} pump signal (amplitude: $V_p$, frequency: $\omega_p$) and applied to the gate electrode. An additional DC voltage (applied using a bias tee) on the gate electrode induces tension in the suspended graphene due to the electrostatic force; this tension provides  tunability of  various modes of the resonator with DC gate voltage (Figure \ref{schematic}c). The gate voltage induced tension in the membrane also allows parametric modulation of the resonator frequency. The device geometry thus enables us to dynamically tune the inter-modal coupling under the action of a detuned pump  using an all electrical configuration.\cite{mahboob2012phonon}

\begin{figure}[p]
\includegraphics[width=.85\columnwidth]{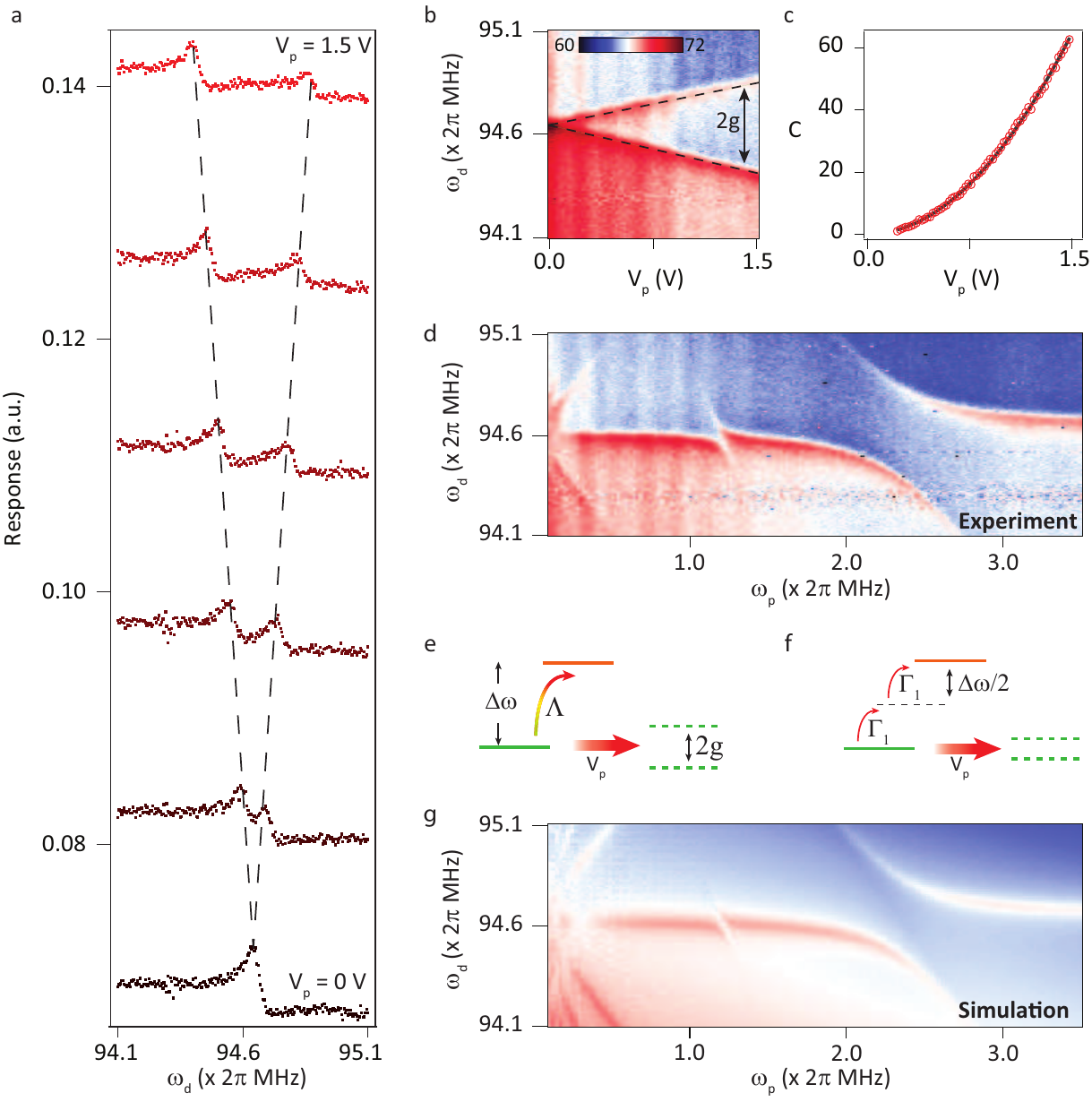}
\caption{\label{powersplit} \textbf{Normal mode splitting and large cooperativity.} \textbf{a},\textbf{b}, With an increase in the red detuned pump amplitude, mode 1 is seen to split into two well resolved peaks with a separation given by $2g$. Constant offset has been added to the response at various pump amplitudes in \textbf{a}. Dotted lines are guides to the eye. Color scale units are $\mu$V. \textbf{c}, Cooperativity of the modes is seen to be as high as 60 at the largest pump amplitudes. The solid line is a quadratic fit of the data (circles) to the equation $C=\alpha V_p^2$. \textbf{d}, Mode 1 response is probed as a function of the pump detuning over a larger frequency range with $V_p=1.5$ V. Apart from normal mode splitting at $\omega_p\approx\Delta\omega$, an additional splitting at $\omega_p\approx\frac{\Delta\omega}{2}$ can be seen, suggesting the onset of higher order inter-modal coupling. \textbf{e},\textbf{f}, Schematics to explain the normal mode splitting and avoided crossing due to higher order coupling.  The coupled modes undergo mode splitting in the presence of a red detuned pump when the coupling rate compensates the individual mechanical losses. The coupled modes further interact near $\omega_p=\Delta\omega/2$ due to the second order coupling. \textbf{g}, The simulated response of mode 1 closely resembles the experimental data showing first and second order coupling. Frequency dependent background was added to the simulated response.
}
\end{figure}

Figure \ref{moresplits}a shows the electrostatic tunability with gate voltage of two coupled modes allowing them to be brought into a region of avoided crossing. We refer to the lower (higher) frequency mode as mode 1 (2) in the remainder of the text. The separation between mode 1 and 2 decreases with decreasing $|V_g^{dc}|$ achieving perfect tuning at $V_g^{dc}=-35$ V (see Supplementary Information for detailed measurements on gate tunability of the coupling). The dynamics of the system can be studied under the influence of a red, $\omega_p \sim (\omega_2-\omega_1)=\Delta\omega$, or blue, $ \omega_p \sim (\omega_1+\omega_2)$, detuned pump signal that parametrically modulates the coupling. This is analogous to optomechanical systems where the parametric coupling of a mechanical mirror to an optical cavity can be manipulated to cool, or amplify the mechanical motion, by using red or blue detuned pumps respectively.\cite{kippenberg2007cavity} In our system the first mode (oscillator) is parametrically coupled to the higher frequency second mode which plays the role of a cavity. This allows for experiments where two coupled modes with a large frequency separation can be used for cooling the low frequency mode, equivalent to optomechanical systems.\cite{mahboob2012phonon} In the Supplementary Information we show strong coupling between modes separated by a frequency ratio of $\sim$ 2, however, we now focus on the dynamics of the two, nearby modes in the presence of a red detuned pump.

By applying $V_g^{dc}=-36$ V on the gate electrode, we tune the modes to $\omega_1=2 \pi\times94.65$ MHz and $\omega_2=2\pi\times96.94$ MHz such that $\Delta\omega=2 \pi\times2.29$ MHz. We measure the response of mode 1 as a function of pump frequency at various pump amplitudes. Here, the response of the mode is detected with a weak driving signal of amplitude $\sim$1 mV at $\omega_d$. Figure \ref{moresplits}c shows the response of mode 1 as a function of the pump frequency at $V_p=0$ V. At zero pump power, mode 1 remains unperturbed. When the pump strength is increased to 1.5 V, avoided crossing in the response of  mode 1 is observed in regions where the pump frequency approaches $\Delta\omega$ (Figure \ref{moresplits}d). This avoided crossing signifies a regime of strong coupling where mixing between the modes gives rise to new eigenmodes in the system.\cite{mahboob2012phonon,okamoto2013coherent,liu2015optical} %This is the well known normal mode splitting in coupled systems with new hybridized eigenmodes.
In the strong coupling regime we expect coherent energy transfer between the two modes. This is demonstrated by the measurements given in the inset of Figure \ref{moresplits}d where the response is simultaneously demodulated at  $\omega_d+\omega_p$ with no driving signal at $\omega_2$. Energy transfer to the second mode is observed as a non-zero signal in the region when $\omega_d+\omega_p\approx\omega_2$.

The amount of splitting characterizes the strength of the inter-modal coupling $g$ and it is tunable with the amplitude of the pump signal. Figure \ref{powersplit}a,b shows the response of mode 1 when pumped at a frequency of $\Delta\omega=2 \pi\times2.29$ MHz. As the pump voltage is increased, the resonance peak of mode 1 can be seen to split into two peaks. At moderate powers this peak splitting is equivalent to an optomechanically induced transparency.\cite{weis2010optomechanically,mahboob2012phonon} As the pump is further increased, response of mode 1 splits into well resolved peaks with the separation given by the coupling rate equal to $2g$. The region of splitting is characterized by the coupling rate of the modes becoming larger than their individual dissipation rates ($\gamma_i$). At the highest pump amplitude of $V_p=1.5$ V the splitting ($2g\approx2\pi \times 450$ kHz) exceeds both $\gamma_1\approx2\pi\times64$~kHz and $\gamma_2\approx2\pi\times51$~kHz. The intermodal coupling can be quantified by a figure of merit cooperativity defined as $C=\frac{4g^2}{\gamma_1 \gamma_2}$.\cite{aspelmeyer_cavity_2014}
Figure \ref{powersplit}c shows the cooperativity as a function of the pump amplitude and is seen to be as high as 60 indicating that many cycles of energy transfer can be achieved between the modes before dissipation to the bath. (Additional data showing strong coupling between modes that have frequency ratio of $\sim$ 2 is included in the Supplementary Information.)

The dynamics of our system can be described using the equations of motion for two coupled vibrational modes given by Okamoto et al.\cite{okamoto2013coherent} as:
\begin{equation}
\resizebox{0.7\columnwidth}{!}{$\label{eom1}
\ddot{x}+\gamma_1\dot{x} + (\omega_1^2+\Gamma_1 \cos(\omega_p t))x+\Lambda \cos(\omega_p t)y=F_1 \cos(\omega_d t + \phi)$}
\end{equation}
\begin{equation}
\resizebox{0.7\columnwidth}{!}{$\label{eom2}
\ddot{y}+\gamma_2\dot{y} + (\omega_2^2+\Gamma_2 \cos(\omega_p t))y+\Lambda \cos(\omega_p t)x=F_2 \cos(\omega_d t + \phi)$}
\end{equation}
where $x$ and $y$ are the displacements of the two modes, $\Gamma_i$ are the parametric drives that modulate the stiffness of the modes, $\Lambda$ is the coefficient of mode coupling, and $F_i$ are the forces at drive frequency $\omega_d$. The terms involving $\Lambda$ are responsible for the transfer of energy between the two modes, whereas the terms having $\Gamma_i$, when combined with $\Lambda$, excites higher order coupling between the modes.  The mode splitting, $2g$, is related to the coupling coefficient, $\Lambda$, by $2g\sim\frac{\Lambda}{2\sqrt{\omega_1\omega_2}}$.\cite{okamoto2013coherent} The equations (\ref{eom1})-(\ref{eom2}) are used to solve the dynamics of the system numerically. The above model captures all the experimentally observed features. The coupling coefficients, $\Lambda$ and $\Gamma_i$, are proportional to the pump amplitude, $V_p$.  The experimentally observed splitting follows a linear trend (Figure \ref{powersplit}a) with increasing pump voltage. This is indicative of the $V_p$ dependence of $\Lambda$.

The terms involving $\Gamma_i$ result in higher order coupling between the modes where a second order coupling is characterized by $\Gamma_i\times\Lambda$. In Figure \ref{powersplit}d, the response of mode 1 over a large range of pump frequency shows the emergence of an additional splitting when $\omega_p\approx\Delta\omega/2$. The simulations and experimental results show very good agreement as seen in Figure \ref{powersplit}g. (Supplementary Information shows detailed measurements and comparison between the first and second order coupling rates.)

\begin{figure}[t]
\includegraphics[width=0.7\columnwidth]{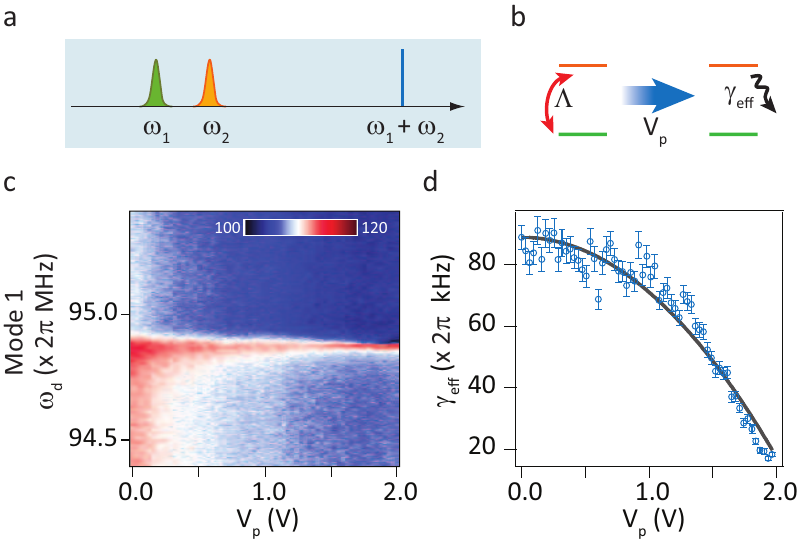}
\caption{\label{bluepump} \textbf{Amplification of motion using blue pump.} \textbf{a}, Illustration of the spectrum shows the two modes along with the blue pump. \textbf{b}, The action of the blue pump is understood as a non-degenerate parametric amplification of the mechanical modes. \textbf{c}, Response of mode 1 as a function of the blue pump amplitude at $V_g^{dc}=44$ V. Color scale units are $\mu$V. The response is seen to narrow with increasing pump amplitude indicative of amplification of the mode. \textbf{d}, The effective dissipation rate of the mode is seen to decrease with the pump amplitude. The data (circles) is fitted to the theoretical response given by the equation $\gamma(V_p)=\gamma_1(1-\beta V_p^2)$.\cite{aspelmeyer_cavity_2014} At $V_g^{dc}=44$ V the initial dissipation rate of mode 1 is $\gamma_1\sim2\pi\times89$ kHz.}
\end{figure}
The nature of the dynamical intermodal coupling can be changed under the action of a blue ($\omega_p\sim\omega_1+\omega_2$) detuned pump. While a red detuned pump swaps the energy between modes, the blue detuned pump amplifies both modes. Figure \ref{bluepump}a shows the frequency spectrum of the modes along with the blue pump ($\omega_1=2\pi\times94.90$ MHz, $\omega_2=2\pi\times102.03$ MHz at $V_g^{dc}=44$ V). In similar optomechanical experiments, a blue detuned pump laser is introduced into a cavity to amplify the motional amplitude of a mechanical resonator where the mechanical loss of the resonator is compensated by the action of the pump, schematically shown in Fig 4b. Here we demonstrate non-degenerate parametric amplification of a mechanical mode by introducing a blue pump signal of frequency $\omega_p=2\pi\times196.93$ MHz on the gate electrode. Figure \ref{bluepump}c shows the response of mode 1 as a function of the pump amplitude. Figure 4d plots the linewidth of  mode 1 with increasing blue detuned pump strength. The narrowing of linewidth can be understood from the non-degenerate parametric down conversion of the pump signal, which compensates the dissipation to the thermal bath, leading to mechanical amplification. The effective dissipation rate can be tuned by a factor of 4 at the highest pump amplitude before reaching the region of instability. The non-degenerate parametric amplification of these high frequency modes could possibly lead to realization of two mode squeezed states at sufficiently low temperatures.\cite{mahboob2014two} The experiments with the blue pump can be treated as a generalized form of parametric amplification of a single mode with a $2\omega$ pump, where the gain is dependent on the phase difference between the drive and pump signals.\cite{rugar1991mechanical} In our system the gate voltage induced tension can be used to parametrically excite a mode at twice its resonant frequency and amplify the mechanical response till a regime of self-oscillations.\cite{barton_photothermal_2012} In the Supplementary Information we show parametric amplification of the mode near $2\pi\times50$ MHz with a gain factor of nearly 3, limited by nonlinearities in the system.\cite{eichler2011parametric,turner1998five}

In summary, by fabricating a graphene resonator with large frequency tunability, we demonstrate tunable strong coupling between mechanical modes using red and blue detuned pumps. We show normal mode splitting under the influence of red detuned pump achieving a cooperativity of 60. This scenario persists for modes with frequencies differing by a factor of 2 suggesting that by reducing the drum diameter to increase the resonant frequencies, one can realize efficient cooling mechanisms using two coupled phonon modes \cite{wilson-rae_theory_2007,marquardt_quantum_2007}. Recent experiments \cite{singh2014optomechanical,weber_coupling_2014,song_graphene_2014} demonstrating large coupling between a graphene drum's electromechanical modes and an electromagnetic cavity opens up the possibility of realizing entanglement of multiple graphene modes mediated by the electromagnetic cavity bus. Furthermore, the high frequencies attainable in this system could help realize vacuum squeezed states of mechanical motion at low temperatures.\cite{mahboob2014two} Considering the suitability of graphene based NEMS devices as sensors the ability to amplify the motional amplitude in the device makes it furthermore attractive for a wide variety of sensing applications.

\section*{Methods}
\subsection*{Device fabrication}
The devices are fabricated on sapphire substrates to suppress effects of parasitic capacitance. A local gate electrode of evaporated Ti/Pt is defined on the sapphire substrate using e-beam lithography (EBL). Chromium layer of 20 nm thickness is evaporated prior to all EBL steps to avoid charging effects. 300 nm of SiO$_2$ is deposited by plasma enhanced chemical vapor deposition followed by selective reactive ion etching above the gate electrode to form the hole that suspends the graphene. Further EBL is carried out to define the source/drain electrodes on top of the SiO$_2$ near the hole region. Cr/Au is evaporated for the contact electrodes. Finally the graphene flake, exfoliated on a PDMS stamp, is located and draped over the source/drain electrodes and the hole. The van der Waal's forces along the edge of the graphene and SiO$_2$ ensures a sealed drum geometry for the resonator. As the flake is not clamped by metal electrodes, free expansion of the graphene flake relative to the substrate and electrodes results in a sag of the membrane when cooled to low temperatures. This leads to a low built-in tension giving rise to large electrostatic tunability.

\subsection*{Simulations}
The coupled equations \ref{eom1}-\ref{eom2} are numerically solved using \textit{Mathematica} with experimentally obtained parameters as input to the simulation. The steady state response is obtained and compared with experiments. Details on the relevant parameters are given in the Supplementary Information.
\subsection*{Acknowledgments}
We thank V. Singh, A. A. Clerk, A. Bhushan, and A. Naik for discussions and comments on the manuscript. We acknowledge financial support from the Department of Atomic Energy, and Department of Science and Technology of the Government of India (Swarnajayanti Fellowship), and ITC-PAC Grant No. FA5209-15-P-0092.
\subsection*{Author Contributions}
J.P.M performed the experiments, simulations and analyzed the data. R.N.P fabricated the devices and contributed to experiments. A.B. contributed to the fabrication and experiments. R.V. provided input for the measurements. J.P.M. and M.M.D co-wrote the manuscript. All authors provided input on the manuscript. M.M.D supervised the project.

%\bibliography{GPC_references_v3}


%%%%%%%%%% Merge with supplemental materials %%%%%%%%%%
\pagebreak
\widetext
\begin{center}
\textbf{\large Supplementary Information - Dynamical strong coupling and parametric amplification in mechanical modes of graphene drums}
\end{center}
%%%%%%%%%% Merge with supplemental materials %%%%%%%%%%
%%%%%%%%%% Prefix a "S" to all equations, figures, tables and reset the counter %%%%%%%%%%
\setcounter{equation}{0}
\setcounter{figure}{0}
\setcounter{table}{0}
\setcounter{page}{1}
\makeatletter
\renewcommand{\theequation}{S\arabic{equation}}
\renewcommand{\thefigure}{S\arabic{figure}}
\renewcommand{\bibnumfmt}[1]{[S#1]}
\renewcommand{\citenumfont}[1]{S#1}
%%%%%%%%%% Prefix a "S" to all equations, figures, tables and reset the counter %%%%%%%%%%

\section{Parametric amplification and self oscillations}
Figure 1 in the main manuscript shows the DC gate voltage tuning of the resonance modes. Frequency of the lower mode is seen to increase with magnitude of the gate voltage with a nearly constant slope of $\sim1$ MHz/V at higher voltages. The monotonic increase is an indication of low pre-tension in the graphene membrane. As the DC voltage is increased the graphene membrane is pulled closer to the gate electrode that gives rise to tension in the membrane thereby increasing the frequency. We exploit this tunability to modulate the spring constant of the resonator parametrically. This is achieved by applying a pump signal at $2\omega$.

\begin{figure}
\includegraphics[width=0.7\columnwidth]{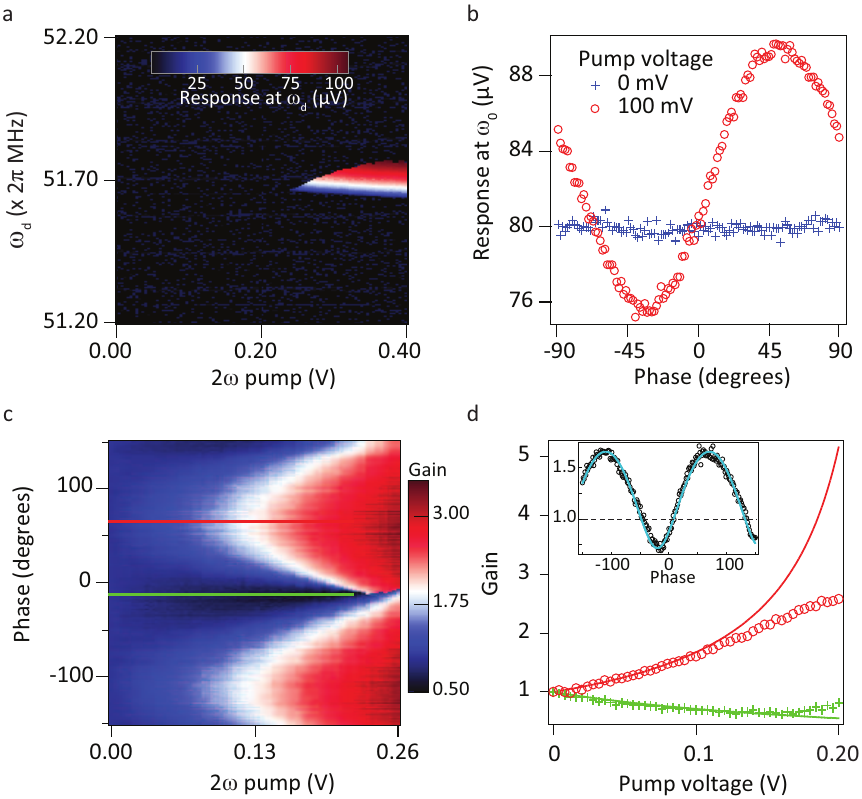}
\caption{\label{sfig:2fpump} \textbf{2$\omega$ pumping induced amplification and self oscillations.} (a) Self oscillation of the resonator is seen beyond critical pump voltage of 248 mV. The response at frequency $\omega$ is plotted as a function of the pump at 2$\omega$ with zero driving signal at $\omega$. (b) Influence of pump voltage at $2\omega_0$ shown as a function of the phase difference between the pump and drive signals. The drive signal amplitude is $\sim1$ mV. (c) Color plot showing the gain of the resonator at $\omega_0$ as a function of phase difference and pump amplitude showing amplification (gain\textgreater1) and de-amplification (gain\textless1). The gain was obtained after subtracting the parasitic background from the signal. (d) Experimentally measured gain along with the theoretical prediction (lines) for two values of phase difference between the drive and pump signals. The experimental data is along the lines of same color from data in (c). The model deviates from the measured gain for pump amplitude above 120 mV. Inset shows the amplification as a function of the phase difference at $V_p=100$ mV along with the fit using equation given in the text.}
\end{figure}

Figure \ref{sfig:2fpump}(a) shows the response of the lower mode at frequency $\omega$ as a function of the pump amplitude at $2\omega$ when the driving force is zero ($\tilde{V}_g=0$ mV). We observe that the graphene enters a regime of self-oscillations beyond a critical pump amplitude ($V_{pc}$) of 248 mV at a resonance frequency of $\omega_0=2\pi\times51.67$ MHz. The tongue shaped response is characteristic of instability in parametrically driven systems.\cite{S_turner1998five} The critical modulation of the spring constant makes the resonator unstable whereby any fluctuations in the system drives it into oscillations. We now apply the pump signal at $2\omega_0$ with an amplitude below the region of instability ($V_p<V_{pc}$) to amplify the mechanical motion. Figure \ref{sfig:2fpump}(b) shows the experimentally obtained response of the resonator driven at $\omega_0$ with pump at $2\omega_0$ as a function of the phase difference between the drive and pump signals. With the pump on we observe a periodic modulation of the otherwise flat signal with phase. The response with pump is seen to go above and below the response at zero pump. This is characteristic of systems with parametric amplification where the response is amplified for certain values of phase difference and de-amplified for others. The gain ($G$) of the oscillator is a function of the phase difference ($\phi$) between the pump and drive and is given by\cite{S_rugar1991mechanical}:
\begin{equation}
\label{gain}
G(\phi)=\left[ \frac{\cos^2(\phi)}{(1+V_p/V_{pc})^2}+\frac{\sin^2(\phi)}{(1-V_p/V_{pc})^2} \right]^{1/2}
\end{equation}
where the gain is measured using the amplitude of oscillation, $z$, as $G=z_{pump\ on}/z_{pump\ off}$. The procedure for obtaining the amplitude of motion from the measured signal is as follows.

The \textit{rf} current ($\tilde{I}$) we measure in our scheme of transduction is related to the amplitude of oscillations ($z$) by:
\begin{equation}
\tilde{I}=i\omega_d C_{p}\tilde{V}_g-i\omega_d \frac{z}{d}C_gV_g^{dc}
\end{equation}
where $C_p$ is the parasitic capacitance between the gate and drain and $C_g$ is the capacitance of the gate at distance $d$ from the membrane. To acquire the response proportional to the amplitude of motion the first term has to be nullified. For this we measure the response at $\omega_d$ keeping $V_g^{dc}=0$ V. This background is then vectorially subtracted from the measured signal to give only the second term. The ratio of measured signals after background correction with the pump on and off then gives the gain of the resonator.

Figure \ref{sfig:2fpump}(c) shows the gain of the resonator as a function of $\phi$ and pump voltage. The modulation in gain is seen to increase smoothly with pump amplitude till the onset of instability near $V_p=V_{pc}$. The maximum amplification of the motional amplitude is seen to be 3, equivalent to 10 dB of gain. This gain, however, is lower than the gain predicted by the above model. Figure \ref{sfig:2fpump}(d) shows the comparison between the measured and theoretically predicted gain. We see that in our resonator the gain stagnates beyond a pump amplitude of $\sim120$ mV whereas equation \ref{gain} predicts a diverging trend. We attribute this deviation to the nonlinear damping in the graphene resonator. The presence of nonlinear terms in the equation of motion can cause saturation of  gain since the dissipation depends on the amplitude of motion.\cite{S_eichler2011parametric}

\section{Equations of motion and simulation}
\begin{figure}
\includegraphics[width=0.75\columnwidth]{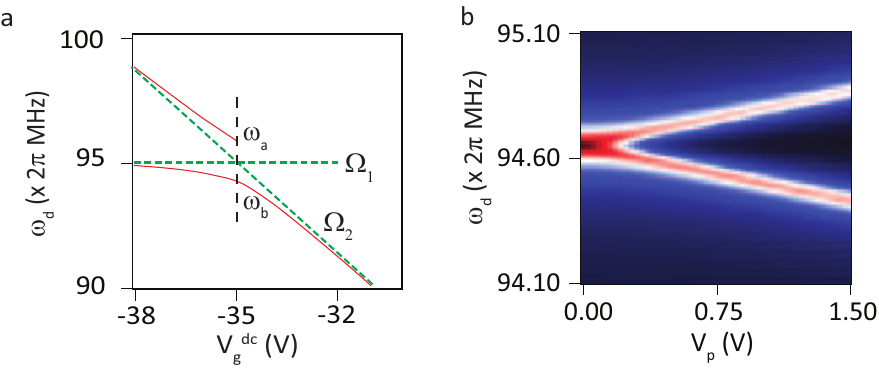}
\caption{\label{sim} \textbf{Numerical solutions of the equations of motion.} (a) Avoided crossing of the two modes with gate voltage. Green dotted lines show the eigenfrequencies ($\Omega_1,\Omega_2$) of the modes. (b) Simulated response of mode 1 as a function of the red pump amplitude. }
\end{figure}
The terms $\Gamma_i$ and $\Lambda$ used in the equations of motion in the main text are given by:\cite{S_okamoto2013coherent}
\begin{equation}
\Gamma_1=\frac{\Gamma}{2}\left(1+\frac{\Omega_1 \delta\Omega}{\sqrt{c^2+\Omega_1^2\delta\Omega^2}}\right)
\end{equation}
\begin{equation}
\Gamma_2=\frac{\Gamma}{2}\left(1-\frac{\Omega_1 \delta\Omega}{\sqrt{c^2+\Omega_1^2\delta\Omega^2}}\right)
\end{equation}
\begin{equation}
\Lambda=\frac{\Gamma c}{2\sqrt{c^2+\Omega_1^2\delta\Omega^2}}
\end{equation}

where $\Omega_i$ are the eigenfrequencies of the two modes without coupling (green dotted lines in Figure \ref{sim}(a)) and $\delta\Omega=\Omega_2-\Omega_1$, $\Gamma=aV_p=\Omega_1\times\delta\Omega/\delta V\times V_p$ is the tunability of the modes, and $c=(\omega^2_a-\omega^2_b)/2$ is the coupling constant ($\omega_a$ and $\omega_b$ are the measured resonant frequencies along the black dotted line in Figure \ref{sim}(a)). For the red pump experiments done at -36 V, $\Omega_1=2\pi\times95.0$ MHz, $\Omega_2=2\pi\times96.2$ MHz, $\omega_a=2\pi\times96.2$ MHz, $\omega_b=2\pi\times94.1$ MHz, and the voltage controlled frequency detuning is $\delta\Omega/\delta V=2\pi \times 1.25$ MHz. Figure \ref{sim} shows the relevant modes along with the simulated normal mode splitting with red pumping. The simulated response is seen to be nearly identical to the experimental result shown in the main text; an increasing pump amplitude increases the splitting between the strongly coupled modes.

\section{Strong coupling: gate tunability and higher order coupling}

\begin{figure}[h]
\includegraphics[width=.75\columnwidth]{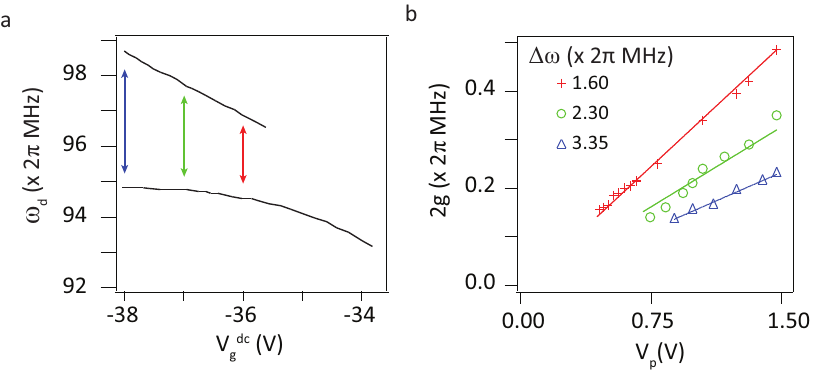}
\caption{\label{sfig:dpumptuning}\textbf{Gate tunability of strong coupling.} (a) The tunability of two modes with gate voltage near the region of avoided crossing is shown. Red pump experiments were done to extract the coupling rate at three different regions of detuning given by the colored lines. The corresponding coupling rates are plotted in (b). Strong coupling between the modes is highest near the region of avoided crossing as seen by the larger $2g$ values. The measured splitting along with a linear fit for three values of dc gate voltage shows the tunability of coupling strength.}
\end{figure}

The coupling between the modes is tunable by the dc gate voltage. The dc gate voltage can be used to bring the modes closer to the region of avoided crossing where the coupling is maximized. Figure \ref{sfig:dpumptuning}(b) shows the splitting of the two modes at different gate voltages. The splitting with pump amplitude (slope of the fit line) is seen to be larger for the dc gate voltage near crossing.

\begin{figure}[p]
\includegraphics[width=0.65\columnwidth]{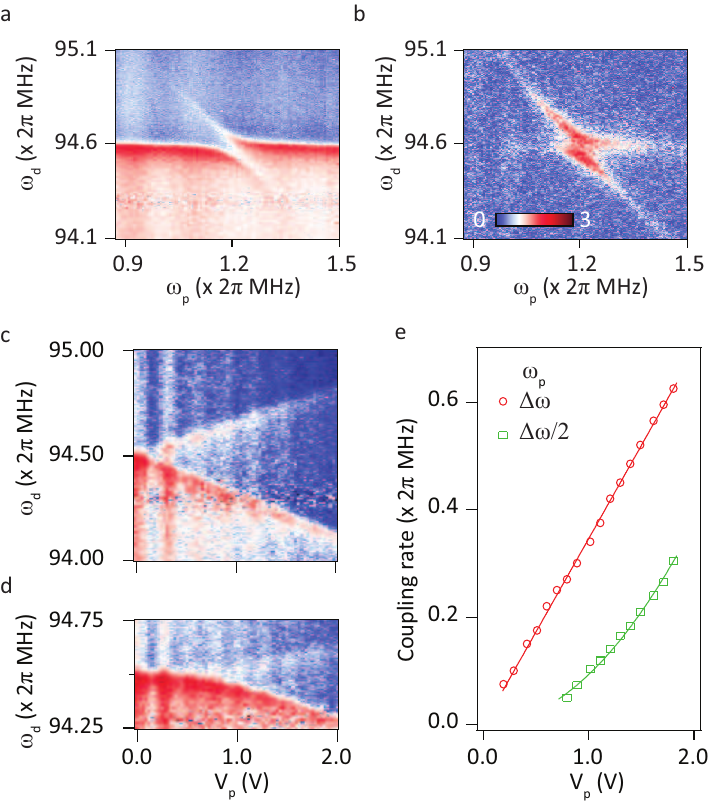}
\caption{\label{sfig:pwrfby2} \textbf{Second order strong coupling and dependence on pump amplitude.} (a) Mode 1 response shows splitting with the pump frequency approaching $\Delta\omega/2$ (b) Simultaneous response detected at the second mode (at $\omega_1+2\omega_p$). Mode 1 response with the pump amplitude for (c) $\omega_p=\Delta\omega$, and (d) $\omega_p=\Delta\omega/2$. The extracted coupling rate is shown in (e) with linear and quadratic fits (red and green lines).}
\end{figure}

The presence of the spring constant modulation ($\Gamma_i$) term in the equation of motion gives rise to additional coupling between the modes when the pump frequency approaches $\Delta\omega/2$. Figure \ref{sfig:pwrfby2}(a) shows a zoomed in measurement of the second order coupling between the modes. Here the relevant parameter of coupling strength is $\Gamma_i\times\Lambda\propto V_p^2$ and the splitting is, hence, expected to follow a quadratic dependence on the pump amplitude as seen in Figure \ref{sfig:pwrfby2}(e).

%\begin{figure}[h]
%\includegraphics[width=0.75\columnwidth]{2ndOrderDpmp-eps-converted-to.pdf}
%\caption{\label{sfig:2ndOrder} Second order strong coupling. (a) Mode 1 response shows splitting with the pump frequency %approaching $\Delta\omega/2$ (b) Simultaneous response detected at the second mode (at $\omega_1+2\omega_p$).}
%\end{figure}

\section{Strong coupling between modes with ratio of frequencies $\sim$ 2}

\begin{figure}[p]
\includegraphics[width=0.6\columnwidth]{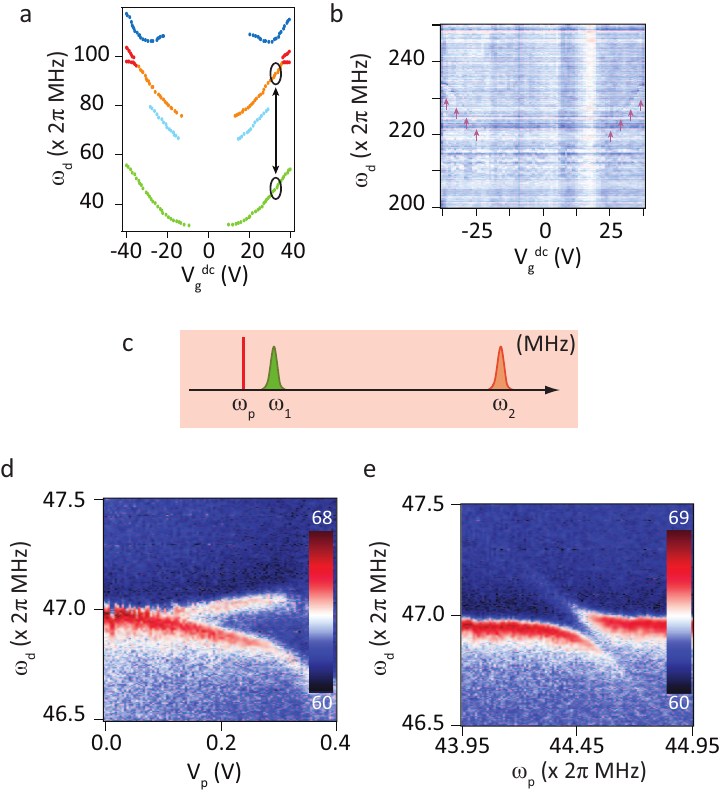}
\caption{\label{diffmodes}\textbf{Normal mode splitting and strong dynamical coupling across modes.} (a) The family of tunable resonance modes is seen as a function of the dc gate voltage. The black circles mark the modes separated by nearly a factor of two on which coupling experiments were done. (b) The highest mode we observe is also seen to have good tunability with the gate voltage. The arrows mark the position of the mode to aid visibility. (c) Schematic showing the spectrum of modes near 47 and 92 MHz along with the red sideband. (d) Response of the lower mode as a function of the amplitude of the pump at a frequency of $\Delta\omega\approx44.7$ MHz shows increasing mode splitting. The response beyond pump amplitude of 0.3 V is not well understood. (e) Normal mode splitting is observed near $\omega_p\approx\Delta\omega$ MHz.}
\end{figure}

Strong coupling can also be achieved between modes of the drum that differ by a factor $\sim2$ in frequency and in general such coupling will exist across modes with higher frequencies. In the main manuscript we show modes that are separated by $\sim$1~MHz. In our drum resonator we have a family of modes with highest gate tunable frequency of $2\pi\times235$ MHz (Figure \ref{diffmodes}(a)-(b)). In order to demonstrate the potential of red-sideband cooling using this coupling across modes, we show dynamical strong coupling between modes whose frequency differs by a factor of 2. Figure \ref{diffmodes}(c)-(e) shows the red sideband experiments for the modes near 47 MHz and 91 MHz. Increasing mode splitting of the mode near 47 MHz is seen with pump amplitude.

\begin{figure}[h]
\includegraphics[width=0.6\columnwidth]{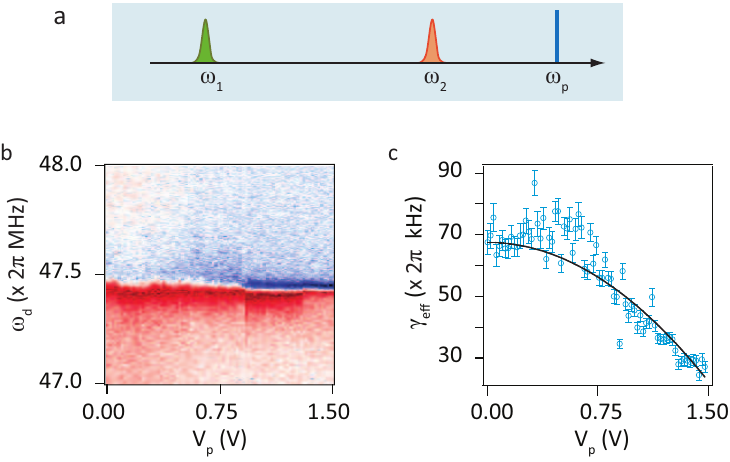}
\caption{\label{spmp2} \textbf{Non-degenerate parametric amplification in well separated modes.} (a) Schematic showing the spectrum of modes near 47 and 92 MHz along with the blue sideband. (b) Response of the 47 MHz mode to a blue pump of frequency $\omega_p=2\pi\times139.28$ MHz. (c) Effective dissipation rate of mode at 47 MHz is seen to decrease with the pump amplitude.}
\end{figure}

Similarly, non-degenerate parametric amplification of the well separated mechanical modes is shown using blue pump experiments. Figure \ref{spmp2} shows the narrowing of the line response of the mode at 47 MHz along with the decreasing effective dissipation rate with blue pump amplitude. The initial dissipation rate was $\gamma_1=2\pi\times68$ kHz.

\end{document}